\documentclass[fleqn,usenatbib,useAMS]{mnras}

\usepackage{graphicx}	% Including figure files
\usepackage{amsmath}	% Advanced maths commands
\usepackage{amssymb}	% Extra maths symbols
\usepackage{multicol}   % Multi-column entries in tables
\usepackage{bm}		    % Bold maths symbols, including upright Greek
\usepackage{pdflscape}	% Landscape pages
\usepackage{todonotes}
\PassOptionsToPackage{hyphens}{url}\usepackage{hyperref}  % "Fix" pdflatex being unable to do multiline hyperlinks

\usepackage{etoolbox}
\makeatletter
\patchcmd\@combinedblfloats{\box\@outputbox}{\unvbox\@outputbox}{}{%
  \errmessage{\noexpand\@combinedblfloats could not be patched}%
}%
\makeatother
\newcommand{\pcmc}{\,cm$^{-3}$~}	% per cm-cubed
\newcommand{\kms}{\,km\,s$^{-1}$~} % kilometres per second
\newcommand{\python}{{\sc Python~}}
\newcommand{\pythonstop}{{\sc Python}}

\usepackage[T1]{fontenc}
\usepackage{ae,aecompl}

\usepackage{newtxtext,newtxmath}
\title[Accretion Disc Winds in Tidal Disruption Events]{Accretion Disc Winds in Tidal Disruption Events: \\Ultraviolet Spectral Lines as Orientation Indicators}

\author[E. J. Parkinson et al.]
{Edward J. Parkinson$^{1}$\thanks{E-mail: \href{mailto:ejp1n17@soton.ac.uk}{e.j.parkinson@soton.ac.uk}}, 
Christian Knigge$^{1}$,
Knox S. Long$^{2,~3}$,
James H. Matthews$^{4}$,\newauthor
Nick Higginbottom$^{1}$,
Stuart A. Sim$^{5},$
and Henrietta A. Hewitt$^{5}$
\\
$^{1}$ Department of Physics and Astronomy, University of Southampton, Southampton, SO17 1BJ, UK\\
$^{2}$Eureka Scientific Inc., 2542 Delmar Avenue, Suite 100, Oakland, CA, 94602-3017, USA\\
$^{3}$Space Telescope Science Institute, 3700 San Martin Drive, Baltimore, MD, 21218, USA\\
$^{4}$Institute of Astronomy, University of Cambridge, Madingley Road, Cambridge CB3 0HA, UK\\
$^{5}$School of Mathematics and Physics, Queen's University Belfast, University Road, Belfast, BT7 1NN, UK\\
}
\date{Accepted XXX. Received YYY; in original form ZZZ}

\pubyear{2020}

\begin{document}
\label{firstpage}
\pagerange{\pageref{firstpage}--\pageref{lastpage}}
\maketitle

%%%%%%%%%%%%%%%% ABSTRACT %%%%%%%%%%%%%%%%%%%%%%%

\begin{abstract}
Some tidal disruption events (TDEs) exhibit blueshifted broad absorption lines (BALs) in their rest-frame ultraviolet (UV) spectra, while others display broad emission lines (BELs). Similar phenomenology is observed in quasars and accreting white dwarfs, where it can be interpreted as an orientation effect associated with line formation in an accretion disc wind.We propose and explore a similar unification scheme for TDEs. We present synthetic UV spectra for disc and wind-hosting TDEs, produced by a state-of-the-art Monte Carlo ionization and radiative transfer code. Our models cover a wide range of disc wind geometries and kinematics. Such winds naturally reproduce both BALs and BELs. In general, sight lines looking into the wind cone preferentially produce BALs, while other orientations preferentially produce BELs. We also study the effect of wind clumping and CNO-processed abundances on the observed spectra. Clumpy winds tend to produce stronger UV emission and absorption lines, because clumping increases both the emission measure and the abundances of the relevant ionic species, the latter by reducing the ionization state of the outflow. The main effect of adopting CNO-processed abundances is a weakening of C~{\sc iv}~1550~\AA~ and an enhancement of N \textsc{v}~1240~\AA~ in the spectra. We conclude that line formation in an accretion disc wind is a promising mechanism for explaining the diverse UV spectra of TDEs. If this is correct, the relative number of BAL and BEL TDEs can be used to estimate the covering factor of the outflow. The models in this work are publicly available online and upon request.
\end{abstract}


%%%%%%%%%%%%%%%%% BODY OF PAPER %%%%%%%%%%%%%%%%%%

\section{Introduction}

    A tidal disruption event (TDE) occurs when an unlucky star passes close to a super-massive black hole (SMBH), and the star's self-gravity is completely overwhelmed by the tidal stresses of the encounter. This results in either the outer layers of the star being stripped away (partial disruption) or the star being completely destroyed \citep[full disruption,][]{Stone2019}. Roughly half of the disrupted stellar debris becomes bound to the SMBH \citep{Rees1988}, accreting this material typically at super-Eddington rates
    and forming a quasi-circular accretion disc \citep[e.g][]{Cannizzo1990, Shiokawa_2015, Hayasaki2016}. This accretion process produces a transient flare whose radiative power is often comparable to the Eddington luminosity, $L_{\text{Edd}}$ \citep{Rees1988}. 
    
    Much of our basic theoretical understanding of TDEs was formulated decades ago \citep[e.g.][]{Hills1975, young_black_1977, hills_stellar_1978, frank_tidal_1978, Rees1988}. However, with recent breakthroughs in transient astronomy and numerical simulation, the established theory has been challenged. For example, characteristic temperatures inferred from the spectral energy distributions of a few recent TDEs are lower than expected \citep[i.e. PS1-10jh;][]{gezari_ultravioletoptical_2012}. Similarly, numerical studies have suggested that the fallback rate of material could in fact be steeper than the fiducial $t^{-5/3}$ power law  \citep[see, e.g., ][]{guillochon_hydrodynamical_2013}, depending on the type of encounter.
    
    Given their extreme luminosities, TDEs are expected to generate significant mass loss in the form of radiation-driven winds. The physical properties of these winds - their geometry, kinematics, mass-loss rates and energetics - remain poorly understood, even though their importance is widely acknowledged. For example, \citet{Miller2015} suggests that the slow temperature evolution and low effective temperatures measured in some TDEs may be due to a wind emanating from the accretion disc, similar to the framework proposed by \citet{laor_line-driven_2014} for AGN. In this picture, a disc wind is responsible for reducing the accretion rate in the inner disc region and thus regulating the effective temperature of the thermal emission. Alternatively, reprocessing of high-frequency photons by an optically thick wind, for example one formed by an accretion disc, could provide an explanation to the origin of optical and UV emission in TDEs \citep[the so-called ``optical-excess'' problem: ][]{Loeb_1997, Guillochon_2014, Metzger2016a, Roth2016, roth_what_2018}. In these reprocessing scenarios, high-frequency X-ray photons, generated at the inner edge of an accretion disc, are absorbed by the optically thick outflow and are reprocessed to lower optical and UV frequencies. This acts to regulate the effective temperature of the thermal emission. Alternative scenarios, in which the optical excess is independent of the accretion process, also exist. In these models, optical and UV radiation is produced by shocks that are formed when streams of bound stellar debris collide during the circularisation process \citep[e.g. ][]{Dai_2015, Shiokawa_2015, Piran2015}. 
    
    Outflows are, in fact, ubiquitous amongst accreting astrophysical systems on all scales, from compact binary systems, such as cataclysmic variables or X-ray binaries, to active galactic nuclei and quasars (QSOs). For all these outflows, there are a number of possible driving mechanisms such as thermal \citep{Begelman1983, Ponti2012}, line/radiation \citep{Castor1975, Lucy1970} and magnetic driving \citep{Blandford1982, Pelletier1992}. One of the clearest outflow signatures in systems containing moderately ionized winds are blue-shifted broad absorption lines (BALs) associated with strong ultraviolet (UV) resonance transitions, such as  Ly$\alpha$ $\lambda 1216$, N {\sc v} $\lambda 1240$, Si {\sc iv} $\lambda 1400$ and C {\sc iv} $\lambda 1550$. These features are seen, for example, in hot stars, accreting white dwarfs (WDs) and QSOs. Recently, blue-shifted BALs have been found in the UV spectra of iPTF15af \citep{Blagorodnova_2019}, iPTF16fnl \citep{Blagorodnova2017} and AT2018zr \citep{Hung2019} at around $\Delta t \approx 50$ days post flare. In another TDE, ASASSN14li \citep{Cenko_2016}, X-ray observations have provided evidence for more highly ionized outflowing gas \citep{miller_flows_2015, kara_ultrafast_2018}. All of these observations imply the existence of powerful outflows in TDEs. 
    
\begin{figure*}
        \centering
        \includegraphics[scale=0.73]{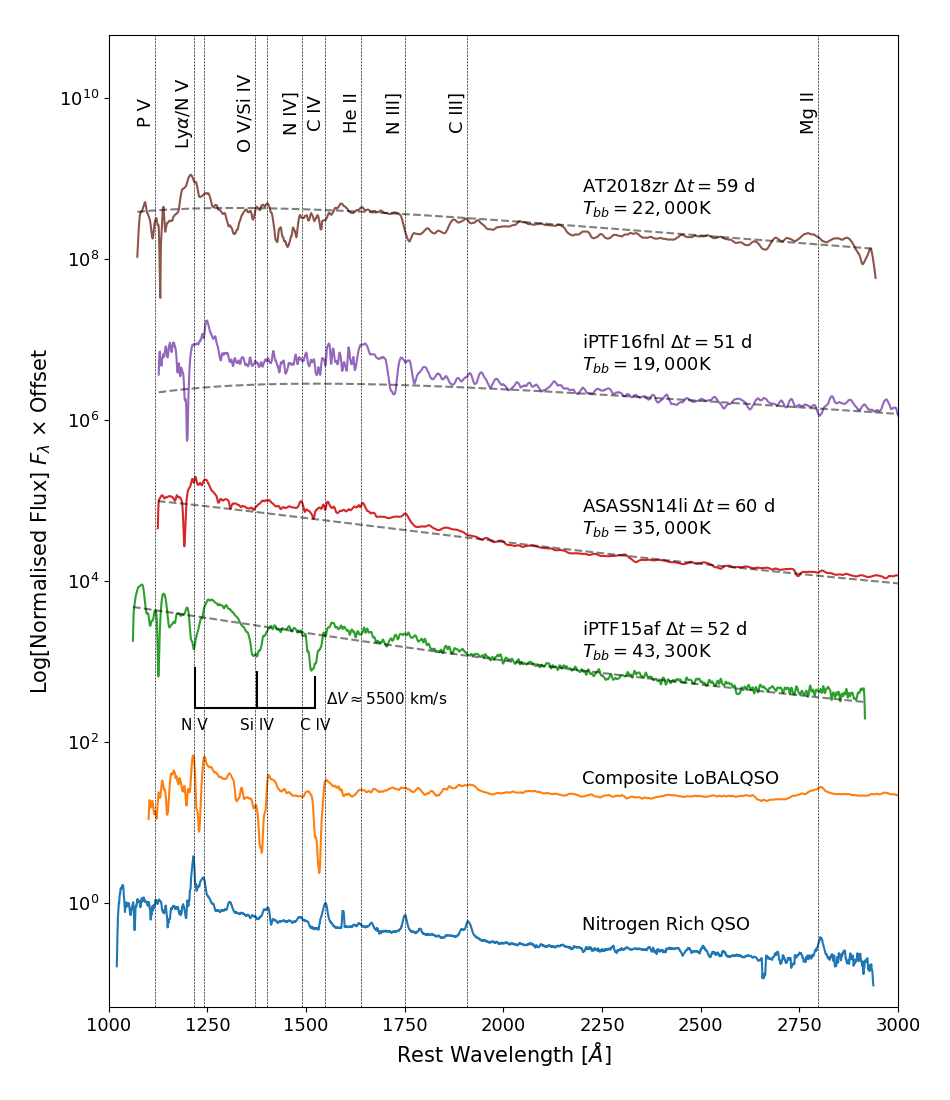}
        \caption{Rest-frame UV spectra of four TDEs; iPTF15af \citep{Blagorodnova_2019};  ASASSN14li \citep{Cenko_2016}; iPTF16fnl \citep{Brown2017, Blagorodnova2017}; and AT2018zr \citep{Hung2019}, observed at similar post-flare epochs. Also included is a composite LoBALQSO spectrum \citep{brotherton_composite_2001} and a rest-frame UV spectrum of the Nitrogen rich QSO SDSS J 164148.19 +223225.22 \citep{Batra2014}. Grey dashed lines mark the continuum for each TDE spectrum, modelled using a simple blackbody of the corresponding temperature. Important line transitions identified are marked at the top. Three important transitions (C {\sc iv}, Si {\sc iv} and N {\sc  v}) are labelled for iPTF15af with approximate blueshift as labelled. All spectra have been smoothed using a boxcar filter.}
        \label{fig: tde_uv_obs}
    \end{figure*}

    Not all TDEs seem to display UV BALs at similar stages of their outburst evolution, however. For example, as illustrated in Figure \ref{fig: tde_uv_obs}, the UV spectrum of ASASSN14li at $\Delta t = 60$ days post flare exhibits predominantly broad emission lines (BELs). This BEL vs BAL dichotomy is reminiscent of Type I QSOs. Most of these display BELs in their UV spectra, but $\simeq 20$ per cent of the population display prominent BAL features \citep[the so-called Broad Absorption Line Quasars (BALQSOs); e.g.][]{Weymann1991, Knigge2008, Dai_2008, Allen2010}. In fact, as illustrated in Figure \ref{fig: tde_uv_obs}, the UV spectra of some TDEs (e.g. iPTF15af and AT2018zr) show a striking similarity to those of BALQSOs. However, the UV spectra of TDE are unique in that the Mg \textsc{ii} $\lambda 2796,2804$ and C \textsc{iii]} $\lambda 1909$ lines, commonly seen in QSO spectra, are either weak or missing entirely \citep[][]{Cenko_2016, Brown2018, Hung2019}. Conversely, TDEs also tend to display strong N \textsc{iii]} $\lambda 1750$ emission, similar to what is seen in the rare Nitrogen-rich QSOs \citep{Osmer1980, Bentz_2004a, Batra2014, Kochanek2016}.
    
    In BALQSOs, line formation in an accretion disc wind can be invoked to explain the broad, blue-shifted UV absorption features \citep{Murray1995, Higginbottom2013}. In fact, simple biconical disc winds can, in principle, produce both the BELs seen in "normal" QSOs and the BALs seen in BALQSOs. More specifically, BELs are always produced, via collisional excitation in the densest parts of the outflow, whereas BALs are seen only if the observer's line of sight to the UV continuum source (the accretion disc) falls within the wind cone \citep[e.g.][]{Shlosman1985, deKool1995, Hamann2013}. The BEL vs BAL dichotomy is therefore an orientation effect in such models \citep[see][]{Murray1995, Elvis2000}. 
    
    Outflows are already a key ingredient in orientation-based unification schemes for TDEs. \citet{Dai2018} have argued that the existence of both X-ray-bright and UV/optically-bright TDEs can be understood as an inclination effect associated with reprocessing in an optically thick disc wind. Outflows have also been invoked as the line-forming region of TDEs, although detailed radiative transfer modelling has so far only been carried out using simple spherical models \citep{Roth2016,roth_what_2018, Dai2018}.
    
    Motivated by these considerations, our goal here is to test whether line formation in disc winds can provide a natural explanation for the BEL vs BAL dichotomy in TDEs. In Section \ref{sec: model_setup}, we describe the calculations and models used in this work. Our results are provided in Section \ref{sec: results}. In Section \ref{sec: discussion} we discuss the implication of these results, as well as the possible effects on them arising from the limitations of our models. Finally, in Section \ref{sec: conclusion}, we summarise our findings.
    
\section{Radiative Transfer and Model Setup} \label{sec: model_setup}

    \subsection{Radiative Transfer and Ionization}
        All calculations for this work were completed using \pythonstop\footnote{\python is a collaborative open-source project. The source code is available at \url{https://github.com/agnwinds/python}.}, a state-of-the-art Monte Carlo ionization and radiative transfer code for moving media that uses the Sobolev approximation \citep[e.g.][]{Sobolev1957, Rybicki1978}. Originally described by \citet{Long2002}, \pythonstop's evolving structure and capabilities have been summarised several times in the literature \citep{Higginbottom2013, Higginbottom2014, Matthews2015, Matthews2016a}. We therefore only provide a brief overview.
        
        \python operates in two stages. The first determines the ionization state, level populations and temperature structure of the modelled outflow. This is done by iteratively flying several populations of Monte Carlo energy quanta (``photon packets'') through a spatially discretised grid. A number of different coordinate systems are supported. The most commonly used are a 2.5D logarithmic polar or logarithmic cylindrical grid. The outflow is discretised over $n_{1} \times n_{2}$ grid cells in these coordinates. Photon packets are generated over a wide wavelength range, sampling the spectral energy distribution (SED) of each radiation source in the model. As these packets interact with the plasma, Monte Carlo estimators that describe the radiation field in each cell are updated. The heating effect of the photon packets is also logged and used to move the temperature towards thermal equilibrium, where heating and cooling are balanced in each cell\footnote{\python includes all significant radiative heating and cooling processes, along with adiabatic cooling due to the expansion of the outflow. Compton heating and inverse Compton cooling are also included, together with Auger ionization and dielectronic recombination \citep[see][]{Higginbottom2013}.}. After each iteration, the revised electron temperature and radiation field estimators are used to update the ionization state of the model. This process is then repeated until the model converges, with heating and cooling balanced throughout the wind.
        
        The second stage of the code produces synthetic spectra. Additional photon packet populations (typically over a narrow wavelength range to give high signal-to-noise) are flown through the \textit{converged} model to generate spectra for a selection of sight lines. 

        \subsubsection{Convergence}
        The convergence of a model in \python is tracked by criteria described by \citet{Long2002}. We consider a grid cell to be converged when i) the electron and radiation temperature have stopped changing between iterations to within 5 per cent, and, ii) when the heating and cooling rates are equal to within 5 per cent. The overall convergence of a model is defined as the fraction of cells that have met these convergence criteria. It is not necessary or expected that all cells converge in a given model. In general, it is cells with poor photon statistics or noisy Monte Carlo estimators that tend to have difficulties converging. These cells are often located near the outer edge of the computational domain and are relatively unimportant in the sense that they do not contribute significantly to the emergent spectra. In this work, models typically have a convergence fraction of $f \succsim 0.85$. However, some of the denser, more optically thick models have convergence fractions of $f \simeq 0.75$.

        \subsubsection{Atomic Data} \label{Section: atomic_data}
        We use the same atomic data as \citet{Long2002}, with the subsequent improvements described by \citet{Higginbottom2013} and \citet{Matthews2015}. Hydrogen and Helium are described with a multi-level model atom and treated using the ``macro-atom'' approach of \citet{Lucy2002, Lucy2003}. By contrast, metals are treated with the original two-level atom formalism adopted in \citet{Long2002}. The resulting hybrid approach is first described by \citet{sim_two-dimensional_2005} and subsequently by \citet{Matthews2015}. By default, \textsc{Python} assumes solar abundances, following \cite{1994A&AS..108..287V}.
        
        \subsubsection{Radiation Sources}
        We include two radiation sources in our model: an accretion disc and the outflow itself. However, only the disc is a \textit{net} source of photons, since all of the emission produced in the outflow is reprocessed disc radiation.
        
        The accretion disc is assumed to be geometrically thin, optically thick and is treated as an ensemble of blackbodies with a standard $\alpha$-disc temperature profile \citep{Shakura1973}. This profile, and hence the emergent SED of the disc, is specified entirely by the mass accretion rate through the disc ($\dot{\text{M}}_{\text{disc}}$) and the central BH mass. The outer radius of the accretion disc is set via a model parameter, $R_{\text{out}} = R_{\text{disc}}$, and the inner radius is set to the innermost stable circular orbit (ISCO) for a non-rotating black hole, $R_{\text{in}} = R_{\text{ISCO}} = 6r_{g}$, where $r_{g} = GM_{\text{BH}}/c^{2}$. We model both fore-shortening and limb-darkening for our disc, resulting in a highly anisotropic radiation field. In reality, accretion discs in TDEs are likely not accurately represented by an $\alpha$-disc. We briefly discuss the limitations of this in Section \ref{sec:limitations}. 
        
        Since the outflow is assumed to be in radiative equilibrium, any energy absorbed by the plasma is reprocessed and reradiated. The reradiation takes place via radiative recombination, as well as free-free and line emission. As these processes depend on the temperature and ionization state of the outflow, the number of photon packets generated due to the outflow is iteratively updated in each cycle of the first (ionization) stage of the code.

        \subsubsection{Microclumping}
        While \python was originally developed to model smooth outflows, real winds are most likely clumpy.  
        Physically, the break-up of a smooth flow into clumps is likely to be the result of instabilities, and several candidates have been identified in various settings \citep[e.g.][]{owocki-unstable,McCourt2018, waters_agn_2019}. 
        Addressing this problem is difficult, however. This is not only because it introduces additional parameters to any model, but also because, if the clouds are sufficiently large, they would need to be resolved in the grid. Consequently, as a first step, \cite{Matthews2016a} implemented a simple approximation known as \textit{microclumping} \citep[e.g.][]{Hamann1998, Hillier1999, Oskinova2008} into \python.
        
        The key assumption in the microclumping approximation is that clumps are optically thin and smaller than all relevant length scales. This is a strong assumption. For example, in the  problem at hand, a photon can interact with a bound-bound transition only in narrow resonance regions, whose width is set by the Sobolev length, 
            \begin{equation}
                l_{\text{S}} = \frac{v_{\text{th}}}{\left| dv/ds \right|}.
            \end{equation}
        Here, $v_{\text{th}}$ is the mean thermal speed and $dv/ds$ is the velocity gradient of the flow. In our models, the Sobolev length is typically $l_{\text{S}} \sim 10^{10} \text{ - } 10^{12}$ cm. On the other hand, line optical depths in a single resonance region can reach $\tau_{max} \sim 10^6$. In order for clumps to remain optically thin even in strong lines, their sizes would therefore need to satisfy $l_{\text{clump}} << l_{\text{S}}/\tau_{max} \sim 10^{4}~\mathrm{cm}$.
        
        The microclumping approximation is nevertheless widely used in stellar wind modelling, at least as a starting point. In the microclumping limit, clumps can be treated simply in terms of a volume filling factor, $f_{v}$. The density of a clump is then multiplied by $D = 1/f_{v}$ relative to the density in the corresponding smooth wind. The inter-clump medium is modelled as a vacuum, under the assumption that it has negligible influence on the emergent spectrum. The opacities, $\kappa$, and emissivities, $\epsilon$, in the model are then given by, 
        \begin{align}
            \kappa &= f_{v} \kappa_{c}(D),\\
            \epsilon &= f_{v} \epsilon_{c}(D),
        \end{align}
        where the subscript $c$ denotes that the quantity is calculated using the enhanced density of the clump. With microclumping, for fixed temperature, processes which scale linearly with density, such as electron scattering, remain unchanged, whereas processes which scale with density squared, such as collisional excitation or recombination, are enhanced.

        The potential impact of clumping is worth considering in TDEs, not least because their spectra resemble those of QSOs (c.f. Figure \ref{fig: tde_uv_obs}). A long-standing challenge for QSO line-driven wind models is that they can easily become overionized when exposed to the intense (E)UV and X-ray radiation field near the central engine \citep[e.g.][]{Proga2002}. Overionization can also prevent the formation of both absorption and emission lines in the UV spectra of disc wind systems, irrespective of driving mechanism \citep[e.g.][]{Higginbottom2013}. Clumping is one natural way to overcome this ``overionization problem'' (e.g. \citealt{Hamann2013, Matthews2016a} -- another is self-shielding, e.g. \citealt{Murray1995,Proga2000, Proga2004}). Overionization can therefore also be expected to be a challenge for spectral models of TDEs. The range of X-ray-to-optical ratios, $L_{\text{X}}/L_{\text{opt}}$, observed in these systems at $\Delta t \approx 60$ d \citep[][]{Wevers2019} is broadly comparable to that seen in QSOs \citep[][]{Steffen2006}. The large dispersion in the $L_{\text{X}}/L_{\text{opt}}$ values seen in both types of systems may be due to geometric effects, highlighting the need for unified models \citep[e.g.][]{Dai2018}.

    \subsection{A Biconical Disc Wind}
    
        \begin{figure}
            \centering
            \includegraphics[scale=0.35]{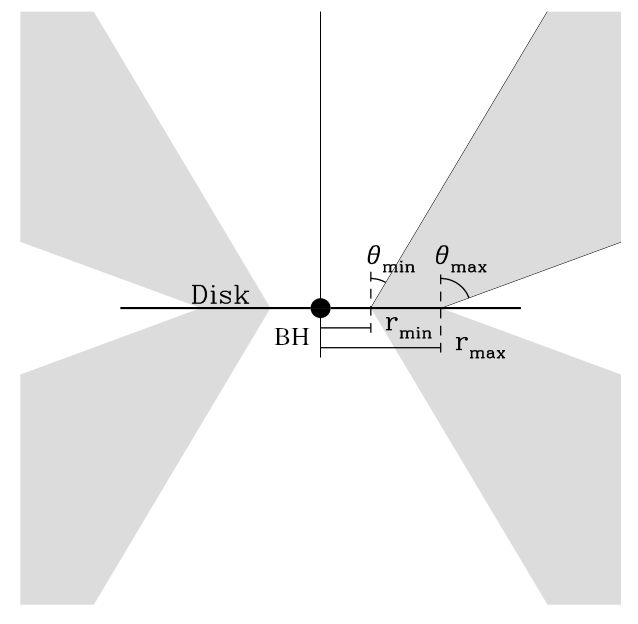}
            \caption{A cartoon showing the basic geometry of the \citet{Shlosman1993} biconical disc wind model.}
            \label{fig: sv93_cartoon}
        \end{figure}
        
        Throughout this work, we examine two distinct wind geometries. Both are based on a kinematic biconical outflow prescription originally designed by \citet{Shlosman1993} and illustrated in Figure \ref{fig: sv93_cartoon}. In this prescription, wind streamlines emerge from disc radii between $r_{\text{min}}$ and $r_{\text{max}}$ at angles relative to the disc normal given by
        \begin{equation}
            \theta(r_{0}) = \theta_{\text{min}} + (\theta_{\text{max}} - \theta_{\text{min}}) x^{\gamma}, 
            \label{eq: sv93_theta}
        \end{equation}
        where,
        \begin{equation}
            x = \frac{r_{0} - r_{\text{min}}}{r_{\text{max}} - r_{\text{min}}}.
            \label{eq: sv93_x}
        \end{equation}
        Here, $r_{0}$ is the launch radius of the streamline, $\theta_{\text{min}}$ and $\theta_{\text{max}}$ are the minimum and maximum opening angles of the wind, and $\gamma$ controls the concentration of streamlines towards either of the two boundaries $r_{\text{min}}$ and $r_{\text{max}}$. 
        
        The launch velocity, $v_{0}$, and terminal velocity, $v_{\infty}$, of a streamline are set to the local sound speed and a multiple of the escape velocity at the streamline footpoint, respectively. The poloidal velocity, $v_{l}$, at a distance $l$ along a streamline, is defined by
        \begin{equation}
            v_{l}(r_{0}) =  v_{0} + (v_{\infty} - v_{0})\left[ \frac{(l / R_{v})^{\alpha}}{(l / R_{v})^{\alpha} + 1} \right],
            \label{eq: sv93_velocity}
        \end{equation}
        where $R_{v}$ is the acceleration length scale, and $\alpha$ controls how rapidly the wind accelerates. The rotational velocity, $v_{\phi}$, is Keplerian at the footpoint of a streamline and  assumed to conserve specific angular momentum as it rises above the disc. At a cylindrical radius $r$,  $v_{\phi}$ is then given by
        \begin{equation}
            v_{\phi}(r) = v_{k} r_{0}/r,
            \label{eq: sv93_rotational_vel}
        \end{equation}
        where $v_{k}$ is the Keplerian velocity at $r_{0}$. The density at any point in the wind is obtained by requiring mass continuity, giving 
        \begin{equation}
            \rho(r, z) = \frac{r_{0}}{r} \frac{dr_{0}}{dr} \frac{\dot{m}^{\prime}}{v_{z}(r, z)}.
        \end{equation}
        In this equation, $\dot{m}^{\prime}$ is the mass-loss rate per unit surface area of the accretion disc, 
        \begin{equation}
            \dot{m}^{\prime} \propto \dot{M}_{\text{wind}} r_{0}^{\lambda} \cos[\theta(r_{0})], 
            \label{eq: sv93_mass_loss_disc_surf}
        \end{equation}
        where $\lambda$ is a parameter that controls where mass is lost from the disc. A value of $\lambda = -2$ results in mass being lost roughly uniformly across the accretion disc surface. 

        \begin{figure*}
            \centering
            \includegraphics[scale=0.6]{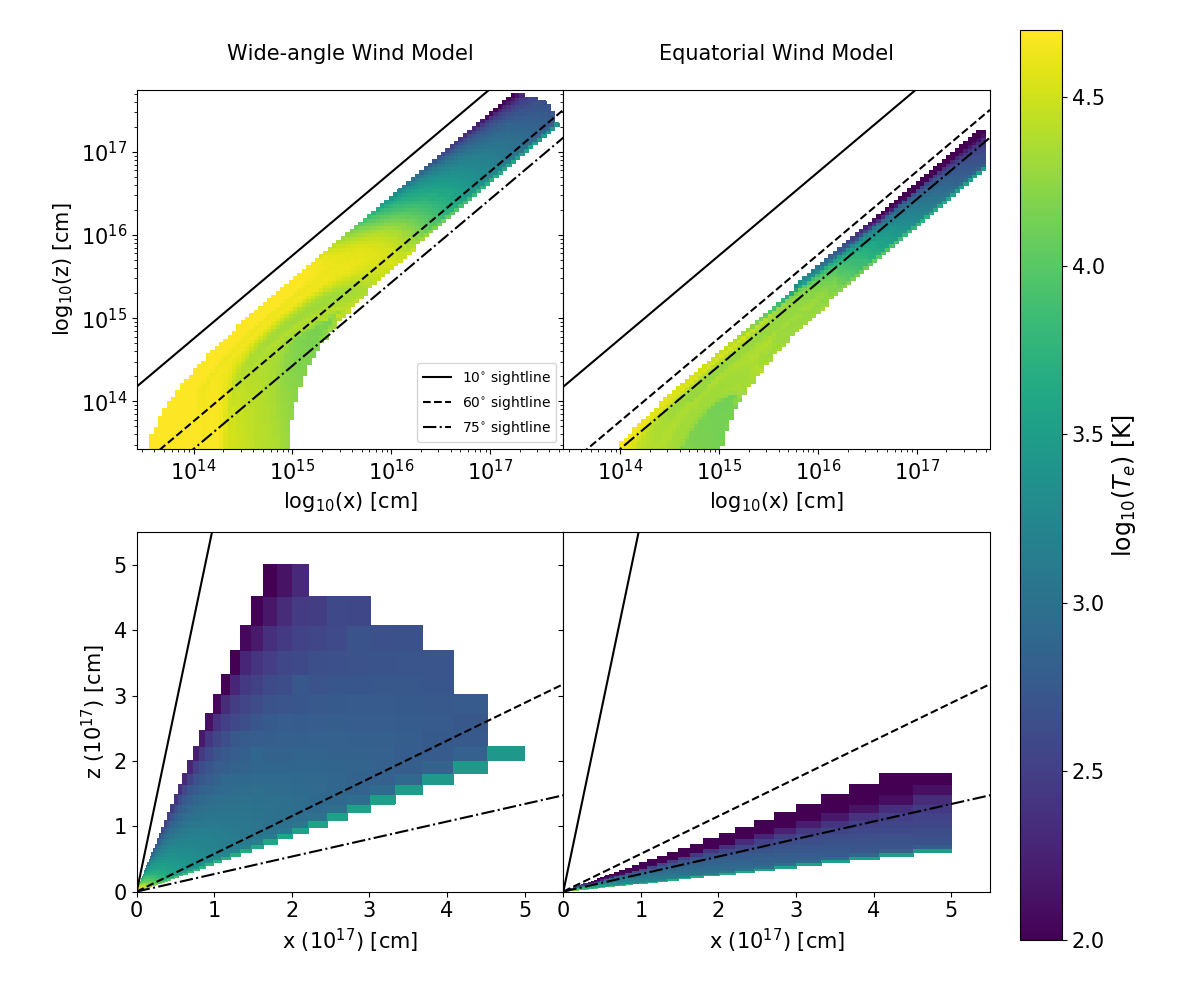}
            \caption{The geometry and temperature structure of our two adopted model geometries, described by the parameters found in Table \ref{tab: model_parameters} and in Section \ref{sec: model_geo_setup}. The colour-map here corresponds to the logarithm of the \textit{converged} electron temperature $T_{e}$. The lines drawn over the wind show sight lines for an observer for inclination angles indicated in the legend. \textit{Top}: the two wind geometries plotted on logarithmic axes. \textit{Bottom}: the same plotted on linear axes.}
            \label{fig: model_geometry_comparision}
        \end{figure*}

    \subsection{Model Setup} \label{sec: model_geo_setup}
    
        We use two sets of model parameters to create (i) a wide-angle and (ii) an equatorial disc wind, inspired by previous efforts to model the outflows from accreting WDs and QSOs, respectively. The difference between these two geometries is illustrated in Figure \ref{fig: model_geometry_comparision}.
        
        We assume a black hole mass of $3 \times 10^{7} \text{M}_{\odot}$ and model an accretion disc using parameters inspired by the characteristic (blackbody) luminosity inferred for iPTF15af \citep{Blagorodnova_2019} to estimate a reasonable accretion rate. More specifically, we assume that the blackbody luminosity of iPTF15af, $L_{\text{BB}} \sim 10^{43} \text{ergs s}^{-1}$, roughly corresponds to the luminosity of the accretion disc, $L_{\text{acc}} = \eta \dot{\text{M}}_{\text{disc}} c^{2}$, for an accretion efficiency of $\eta = 0.1$. We thus calculate an estimate of the accretion rate to be roughly $\dot{\text{M}}_{\text{disc}} \simeq 2 \times 10^{-2} \text{M}_{\odot} \text{yr}^{-1}$, corresponding to an Eddington accretion limit of $0.02~\dot{\text{M}}_{\text{Edd}}$ or $0.01~L_{\text{Edd}}$. The outer radius of the disc for both models is set to be $R_{\text{disc}} = 10^{15}$ cm to roughly match the continuum level of iPTF15af. As shown later in Section \ref{sec: BELBAL}, BALs are produced by both models, but for different viewing angles. We therefore adjusted the accretion rates of each model so that the predicted model flux for a BAL-forming sight line roughly matches the flux level of iPTF15af at $d = 350$ Mpc. For the equatorial model, we found that the previously calculated accretion rate already matched well with observations. However, the wide-angle model required a slightly lower mass accretion rate of $\dot{\text{M}}_{\text{disc}} = 10^{-2} \text{M}_{\odot} \text{yr}^{-1}$. This difference is mainly due to foreshortening: BALs are observed for more face-on inclinations in the wide-angle model.
        
        For both models, we initially assumed a mass-loss rate of $\dot{\text{M}}_{\text{wind}} = 0.1\, \dot{\text{M}}_{\text{disc}}$; based on previous experience of modelling disc winds in accreting WD and QSO systems. We conducted an initial coarse parameter search, focusing on parameters which can impact line formation, i.e. the velocity law and mass-loss rate of the wind, consequently finding that a larger mass-loss rate of $\dot{\text{M}}_{\text{wind}} = \dot{\text{M}}_{\text{disc}}$ resulted in stronger line formation for the wide-angle model. Our final parameter choices for the two wind models are presented in Table \ref{tab: model_parameters}. We conducted a finer parameter search around these values in Table \ref{tab: model_parameters}, finding that the ionization state and emergent spectra are fairly insensitive to moderate changes in these values.
    
        In both models, the disc wind emanates from the entire surface of the accretion disc. We choose values of $\gamma = 1$ for uniform streamline spacing and $\lambda = 0$ corresponding to uniform mass loss across the disc surface. We use a logarithmic cylindrical grid with a greater concentration of cells at smaller radius, where most of the line formation is expected to occur. Several resolution tests were conducted. For grids with a spatial resolution lower than $50 \times 50$ cells, we found that the line-forming region was inadequately spatially resolved. For resolutions higher than $75 \times 75$ cells, we found only small changes to  the emergent spectra. We use $100 \times 100$ cells for our final calculations, ensuring we adequately resolve the line-forming regions. We adopt a maximum wind radius of $R_{\text{wind}} = 5 \times 10^{17}$ cm to ensure our computational domain is large enough to not to affect the results. We ran several models to test this, varying the maximum wind radius by factors of a few, finding little effect on the emergent spectra.

        \begin{table}
            \centering
            \begin{tabular}{cccc}
                \hline
                Parameters & Wide-angle & Equatorial & \\
                \hline
                $r_{\text{min}}$ & $2.65 \times 10^{13}$ & $2.65 \times 10^{13}$ & cm \\
                $r_{\text{max}}$ & $7.95 \times 10^{14}$ & $7.95 \times 10^{14}$ & cm \\
                $\alpha$ & 1.5 & 1.0 &  - \\
                $v_{\infty}$ & 1.0 & 1.0 &  $v_{\text{esc}}$ \\
                $R_{v}$ & $5 \times 10^{16}$ &  $5 \times 10^{16}$ & cm \\
                $\theta_{\text{min}}$ & 20 & 70 & $^{\circ}$ \\
                $\theta_{\text{max}}$ & 65 & 82 & $^{\circ}$ \\
                $\dot{M}_{\text{wind}}$ & $10^{-2}$ & $2 \times 10^{-3}$ & $\text{M}_{\odot} \text{yr}^{-1}$ \\
                $\dot{M}_{\text{disc}}$ & $10^{-2}$ & $2 \times 10^{-2}$ & $\text{M}_{\odot} \text{yr}^{-1}$ \\
                \hline
            \end{tabular}
            \caption{Parameters and their respective values adopted for our two wind geometries presented in Figure \ref{fig: model_geometry_comparision}.}
            \label{tab: model_parameters}
        \end{table}

\section{Results} \label{sec: results}

    In the following sections, we present the results of our simulations. First, in Section \ref{sec: BELBAL}, we present a fiducial clumpy wide-angle outflow using solar abundances. We discuss the physical properties of the outflow and examine its synthetic spectra in the context of how an accretion disc wind could be responsible for the observed BEL vs BAL dichotomy. We then examine the effects of wind geometry in Section \ref{sec: geo_comp}, followed by how microclumping  and abundance variations impact the emergent spectrum in Section \ref{sec: clump_results} and Section \ref{sec:abund_results} respectively. The synthetic spectra, and their respective \python parameter files, described in the following sections are available upon request or are publicly available online at \url{https://github.com/saultyevil/tde_uv_disc_winds_2020}.
    
    \subsection{A Fiducial Wide-Angle Disc Wind} \label{sec: BELBAL}
        
        We now describe a clumpy wide-angle disc wind with solar abundances and clumping factor $f_{v} = 0.1$ in the context of the formation of BALs and BELs. Based on our limited exploration of parameter space (see Sections \ref{sec: geo_comp}, \ref{sec: clump_results}, \ref{sec:abund_results} below), we use this as our %"best-bet"
        fiducial disc wind model.
    
        In Figure \ref{fig: best_model_wind}, we present a selection of physical parameters for the wide-angle wind on a log-log distance scaling. In Figure \ref{fig: best_model_data_comparison_clump}, we show the synthetic UV spectra produced by this model compared to the spectra of a BAL TDE (iPTF15af) and a BEL TDE (ASASSN14li). For our BAL model, we adopt an inclination of $i = 60^{\circ}$, i.e. a sight line which looks into the wind cone. To represent BEL models, we show spectra for both face-on (i = $10^{\circ}$) and edge-on ($i = 75^{\circ}$) sight lines, both of which lie outside the wind cone.
    
        \subsubsection{The Physical Properties}
        
        \begin{figure*}
            \centering
            \includegraphics[scale=0.53]{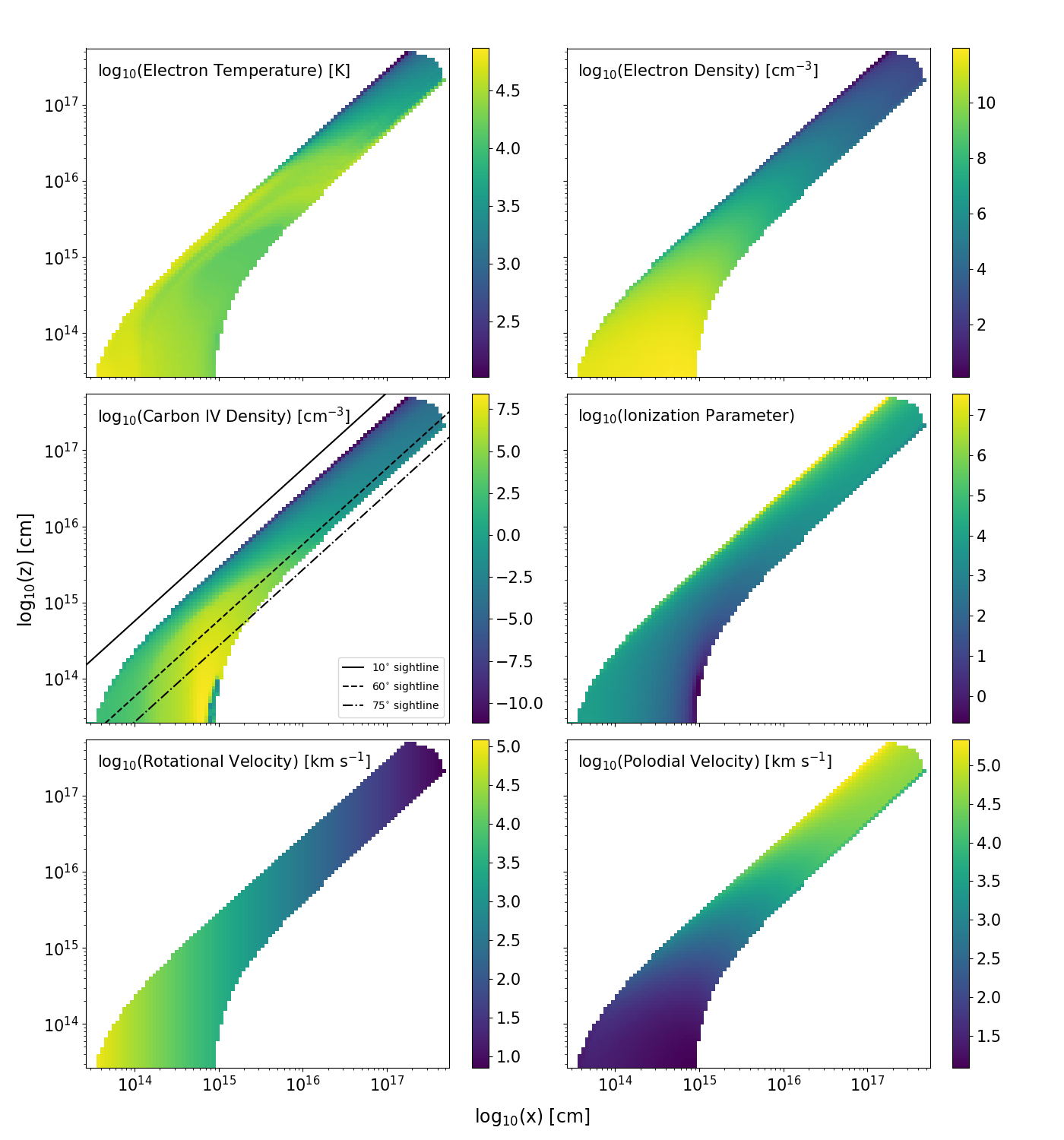}
            \caption{Contour plots showing a selection of physical properties for our clumpy wide-angle disc wind, with clumping factor $f_{v} = 0.1$. Only the $x$-$z$ plane is shown where the wind is rotationally symmetric about the $z$-axis. The lines drawn over the wind show sight lines for an observer for the inclination angles indicated in the legend. The spatial scales and colour maps are logarithmic. \textit{Top left:} the electron temperature. \textit{Top right:} the electron number density. \textit{Middle left:} the C \textsc{iv} ion density. \textit{Middle right:} the ionization parameter. \textit{Bottom left:} the rotational velocity. \textit{Bottom right:} the polodial velocity.}
            \label{fig: best_model_wind}
        \end{figure*}
        
        Figure \ref{fig: best_model_wind} shows a selection of physical parameters for the converged wide-angle wind. Considering first the electron number density of the wind, we see that the base has a high density with $n_{e} \sim 10^{11}~\text{cm}^{-3}$. The electron density declines gradually with radius due to the expansion and acceleration of the wind, resulting in line formation processes which scale with density decreasing in the outer wind regions. The velocity at the base of the wind is dominated by rotation, where it is effectively Keplerian. Here, the velocity is $v_{\phi} \sim 10^{5}$ \kms at the inner disc edge and $v_{\phi} \sim 10^{4}$ \kms at the outer edge. Further out in the wind, the rotational velocity decreases linearly due to the wind material conserving specific angular momentum. The polodial velocity of the outflow,  however, increases with distance, as determined by our velocity law exponent. At the base of the wind, $v_{l} \sim 10$ \kms, but $v_{l} \sim 10^{5}$ \kms ($0.3c$) is reached near the outer edge. Doppler broadening of emission and absorption lines is dominated by rotation near the disc-plane and by the poloidal outflow velocity further out. 
        
        The hottest region of the wind is at the base, where it is exposed directly to the radiation field from the accretion disc. Here, the electron temperature is $T_{e} \sim 3 \times 10^{4}$ K. At larger radii, near the outer edges, where adiabatic cooling dominates, the wind is cooler with $T_{e} \sim 10^{3}$ K and can be as low as $T_{e} \sim 10^{2}$ K where the density of the wind is lowest. In these regions, the wind is illuminated by the reprocessed disc SED. The flux of high-frequency photons is reduced significantly in these regions, resulting in less energy being dumped into the wind and subsequently less heating. In reality, dust and/or molecules could potentially form in this region. However, since \python does not include any dust or molecular physics, our treatment of these regions is highly approximate but should not contribute significantly to the formation of the UV features we are interested in. It is also worth noting that the ionization state of these cells could be \textit{frozen-in}, i.e. set by the advection of the physical properties and ionization state of the outflow at smaller radii  \citep[see, e.g.][]{Owocki1983}. In \python, the ionization state is always computed based on the local properties of the radiation field and grid cell, i.e. this kind of freezing-in is not modelled. However, at the large radii where this is relevant, the ionization state is roughly constant along flow lines in our models.
        
        It is possible to characterise the ionization state of the wind by using an ionization parameter, $U_{\text{H}}$, given by, 
        \begin{equation}
            U_{\text{H}} = \frac{4\pi}{n_{\text{H}}c} \int_{13.6 \frac{\text{eV}}{h}}^{\infty} \frac{J_{\nu}}{h\nu}~d\nu,
        \end{equation}
        where $\nu$ denotes frequency, $n_{\text{H}}$ is the number density of Hydrogen, $c$ is the speed of light, $h$ is Planck's constant and $J_{\nu}$ is the monochromatic mean intensity. The ionization parameter measures the ratio of the number density of Hydrogen ionizing photons to the local matter density. For values of $U_{\text{H}} > 1$, Hydrogen is ionized making it a useful predictor of the global ionization state. Throughout most of the wind, $U_{\text{H}}$ is roughly constant with $U_{\text{H}} \sim 10^{4}$. As $U_{\text{H}} > 1$, this indicates that Hydrogen is ionized throughout the wind and suggests that $n_{\text{H}} \simeq n_{e}$. However, the ``top'' of the wind is highly ionized with $U_{\text{H}} \sim 10^{7}$. The UV absorption lines form along sightlines where the relevant ion density is large, i.e. where the C \textsc{iv} ion density $n_{\text{C IV}} \gtrsim 10^{6}$ \pcmc, approximately traced along the $60^{\circ}$ sightline (middle left panel of Figure \ref{fig: best_model_wind}). The UV emission lines, on the other hand, preferentially form in the base of the wind, where the electron density is high.
        
        \subsubsection{The Emergent Spectrum}
        
        \begin{figure*}
            \centering
            \includegraphics[scale=0.71]{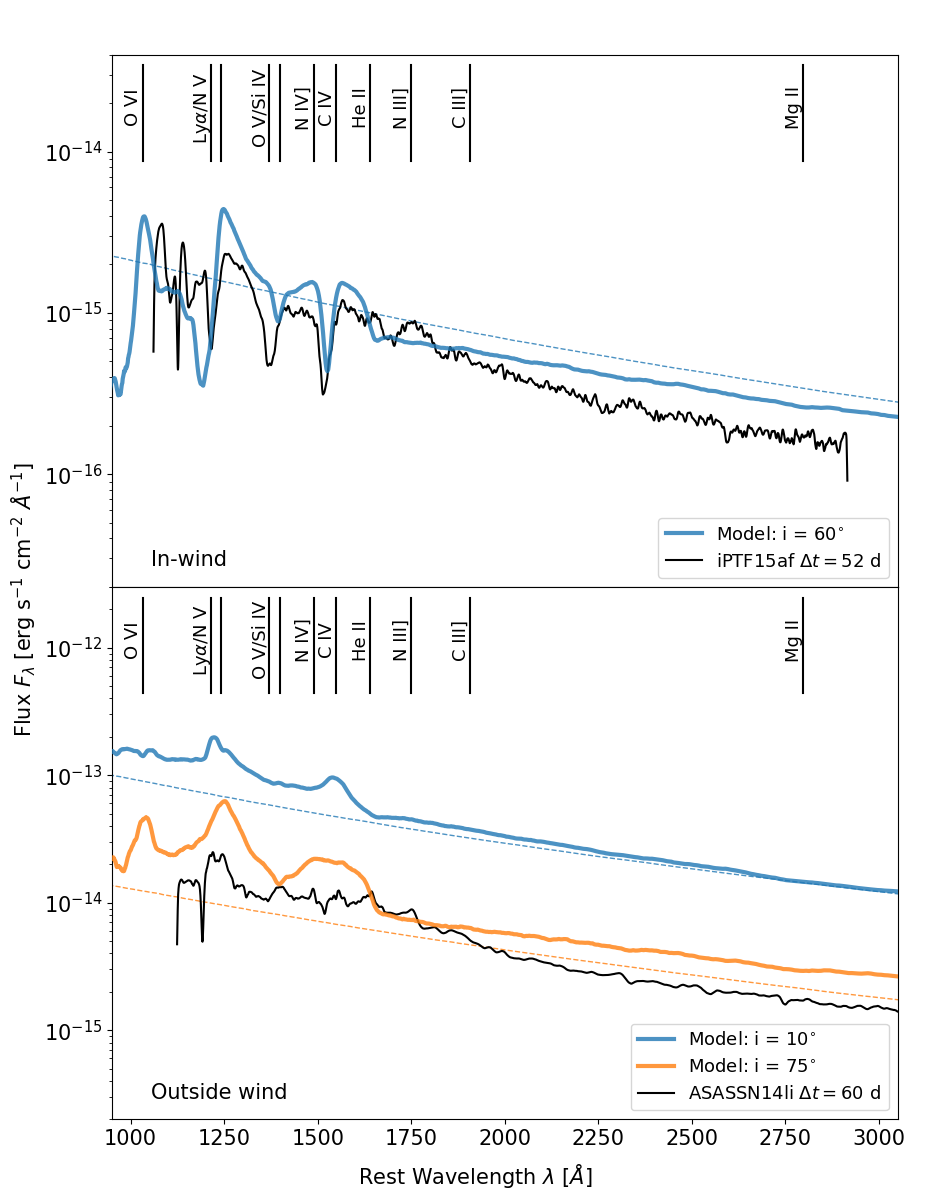}
            \caption{Synthetic UV spectra for our clumpy wide-angle disc wind with a clumping factor $f_{v} = 0.1$ for three sight line angles. Over-plotted is both a characteristic BAL TDE (iPTF15af) and BEL TDE (ASASSN14li) for comparison to the model. Marked by a dashed line is the accretion disc SED for the clumpy wide-angle model. The flux of the model has been scaled to the distance of the relevant TDE which it is being compared to. Important line transitions in the spectra have been marked at the top of each panel. \textit{Top panel}: an observer with a line of sight through the wind cone which observes BALs. \textit{Bottom panel}: an observer with a line of sights not looking through the wind cone, observing BELs.}
            \label{fig: best_model_data_comparison_clump}
        \end{figure*}
        
        The top panel of Figure~\ref{fig: best_model_data_comparison_clump} shows the BAL case. The model produces broad absorption in the strong resonance lines of C \textsc{iv}, Si \textsc{iv} and N \textsc{v}. Our wide-angle BAL model is clearly far from a perfect match to iPTF15af. However, it is encouraging that it produces BALs in essentially the correct set of transitions, for a UV continuum that reasonably matches observations in both shape and normalisation. 
        
        One interesting feature of the BAL model is that the profiles of C \textsc{iv} and N \textsc{v} are significantly blue-shifted, whereas the Si \textsc{iv} absorption feature is centred close to the rest wavelength of the rest transition. This reflects the ionization state of the model. The dominant ionization stage of Silicon throughout most of the outflow is actually Si~\textsc{v}. As a result, Si \textsc{iv} is typically formed only in the dense, low velocity base of the wind resulting in a small blueshifting of the Si \textsc{iv} absorption line. The model spectrum also features broad and blueshifted O \textsc{vi}~$\lambda 1034$, representing an even higher ionization potential than N \textsc{v}.
        
        The bottom panel of Figure \ref{fig: best_model_data_comparison_clump} shows the BEL spectra produced by our clumpy wide-angle model for face-on and edge-on sight lines. At $i = 10^{\circ}$, the model produces strong C \textsc{iv}, N \textsc{v} and  Ly$\alpha$ emission lines. The same transitions are also seen in emission at $i = 75^{\circ}$, but the lines are now so broad that N \textsc{v} and  Ly$\alpha$ are blended. Neither spectrum produces Si \textsc{iv} emission, again reflecting that Si \textsc{v} dominates throughout most of the outflow volume. However, a weak absorption trough associated with Si \textsc{iv} can be identified in the edge-on spectrum, which cuts through the dense, low velocity base of the wind where Si \textsc{iv} ions are preferentially found. The BELs seen in ASASSN14li are considerably narrower than those produced by the wide-angle model for $i=75^{\circ}$, favouring the association of BEL TDEs with a face-on system.
        
        \subsection{Wide-angle vs Equatorial Winds} \label{sec: geo_comp}
        
        The spectra produced by both wide-angle and equatorial wind models are shown in Figure~\ref{fig: model_comparison_clump} (for smooth vs clumpy outflows) and Figure~\ref{fig: model_comparison_abund} (for solar vs CNO-processed abundances). The dependence of the spectra on clumpiness and abundances will be discussed in Sections~\ref{sec: clump_results} and \ref{sec:abund_results}, respectively. Here, we briefly comment first on the basic differences between the spectra produced by the two types of outflow geometry. The implications of these differences are discussed in more detail in Section~\ref{sec: discussion-geo}.

        In principle, both wide-angle and equatorial winds are able to produce BALs and BELs. In both geometries, BALs are observed preferentially for sight lines looking into the wind cone. However, wide-angle flows produce BELs for both low and high inclinations, whereas our equatorial model only produces BELs at low inclinations. Perhaps more importantly, in terms of line-to-continuum contrast, the equatorial model only produces very weak emission lines. This is partly due to the anisotropic radiation pattern produced by the disc in our model. As a result, the continuum level is much higher for the face-on orientations that can produce BELs in the equatorial wind model. The line emission produced by the wind is always more isotropic than the disc continuum, so it is harder to achieve high line-to-continuum contrasts in equatorial models. As discussed further in Section~\ref{sec: discussion}, it is actually not clear if real TDE discs emit this anisotropically. For a more isotropic continuum source, the \textit{emission} line-to-continuum contrast will be considerably more uniform across all orientations. However, \textit{absorption} features would still be formed preferentially for sight lines looking into the wind cone.
        
    \subsection{Clumping} \label{sec: clump_results}

        \begin{figure*}
            \centering
            \includegraphics[scale=0.53]{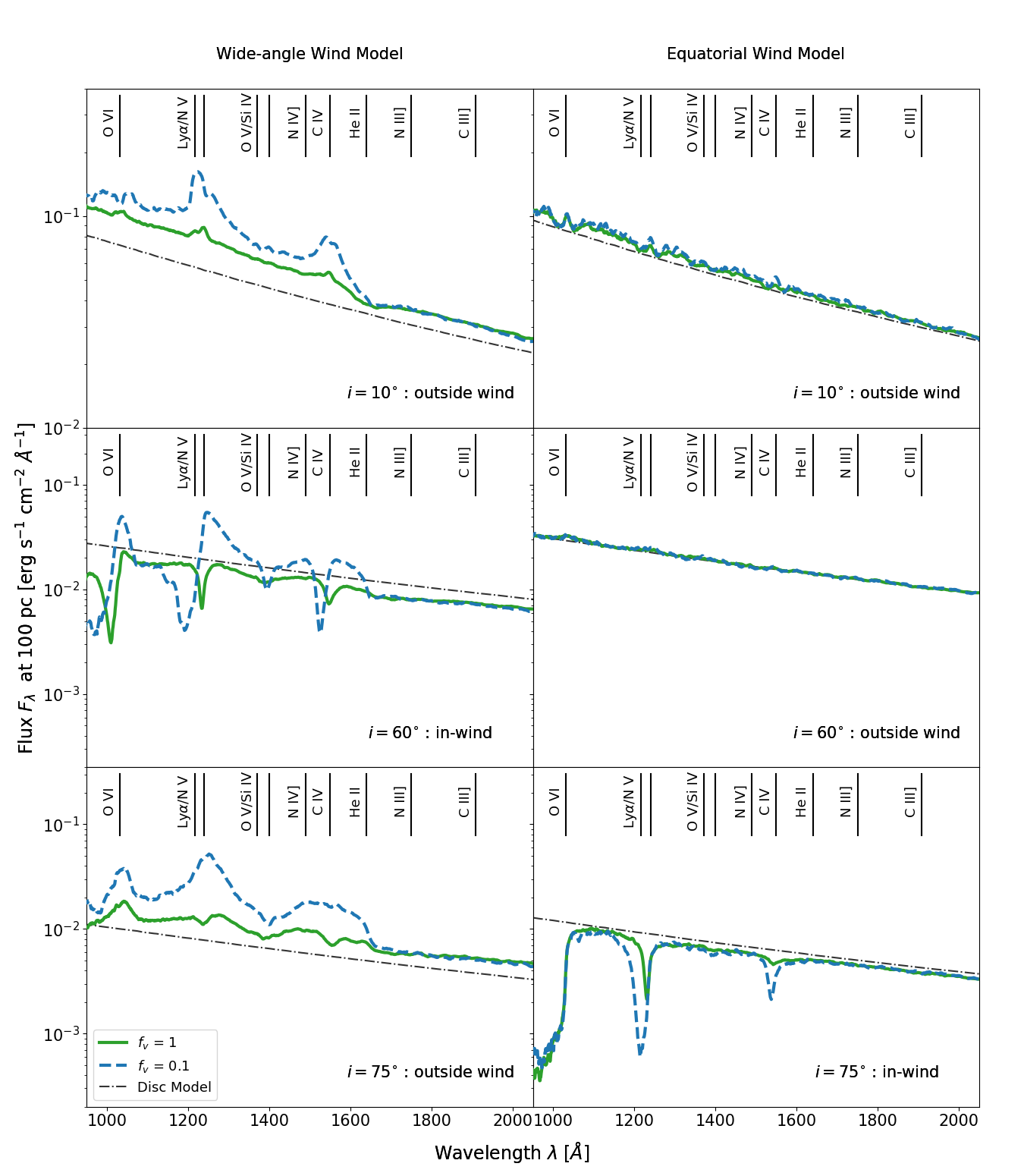}
            \caption{Synthetic UV spectra generated for a smooth and clumped wind for our two wind geometries at characteristic low, medium and high inclination angles. In the figure smooth wind spectra are displayed as solid lines, whereas clumpy wind spectra are shown with dashed lines. The underlying accretion disc SED (dot-dash lines) for each model are also plotted.}
            \label{fig: model_comparison_clump}
        \end{figure*}

        Figure \ref{fig: model_comparison_clump} illustrates the effect of clumping on the emergent spectra. More specifically, we compare here the spectra produced by smooth winds ($f_{v} = 1$) to those produced by clumpy winds with a clump to inter-clump density ratio of 10 (i.e. $f_{v} = 0.1$). 
    
        The main effect of clumping is to strengthen the UV resonance transitions. This happens because clumping increases the density of the line-forming gas and hence lowers its ionization state. In the wind models presented here, which tend to be somewhat overionized, this increases the abundance of the ionic species responsible for the UV lines. For the ``in-wind'' (BAL) spectra, this manifests as deeper and broader absorption troughs. The BALs are also more highly blue-shifted, as the abundance of the relevant ions in the high velocity regions of the wind is increased. Emission lines are additionally strengthened in clumpy winds, because collisional excitation scales as the \textit{square} of the density in the line-forming region. 
        
        The presence of clumps is particularly important for the formation of Si~{\sc iv}. As noted before, Si~{\sc v} tends to be the dominant ionization stage of Silicon in our models, but clumpy winds contain enough Si~{\sc iv} to produce BAL features for sight lines looking into the outflow. This feature is not generally observed in our smooth wind models. 
        
    \subsection{Abundances} \label{sec:abund_results}
    
        \begin{figure*}
            \centering
            \includegraphics[scale=0.52]{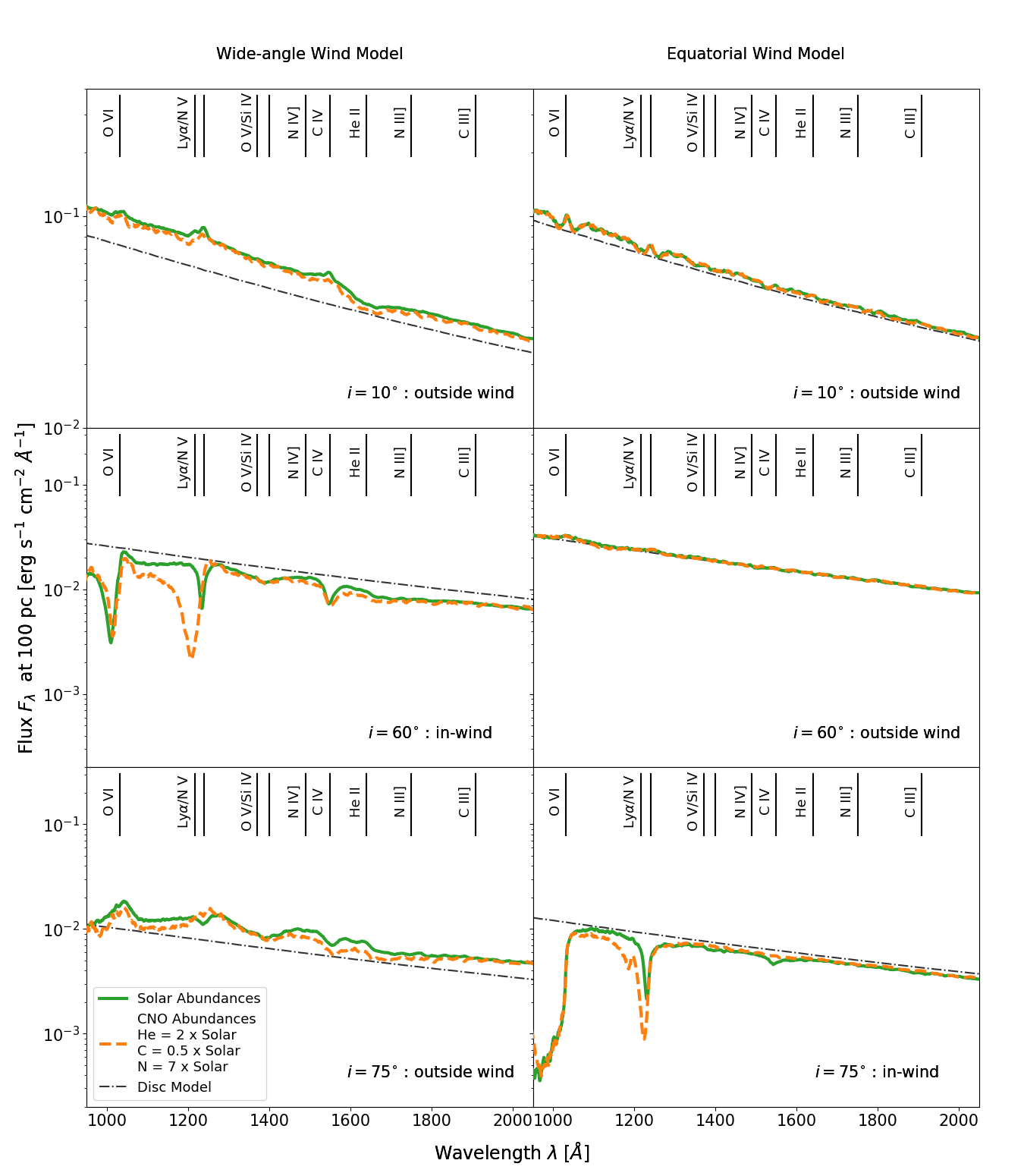}
            \caption{Synthetic UV spectra generated for smooth winds for our two wind geometries for both Solar and CNO-processed abundances at characteristic low, medium and high inclination angles. In the figure Solar wind spectra are displayed as solid lines, whereas CNO-processed wind spectra are shown with dashed lines. The underlying accretion disc SED (dot-dash lines) for each model are also plotted.}
            \label{fig: model_comparison_abund}
        \end{figure*}
    
\begin{figure*}
            \centering
            \includegraphics[scale=0.52]{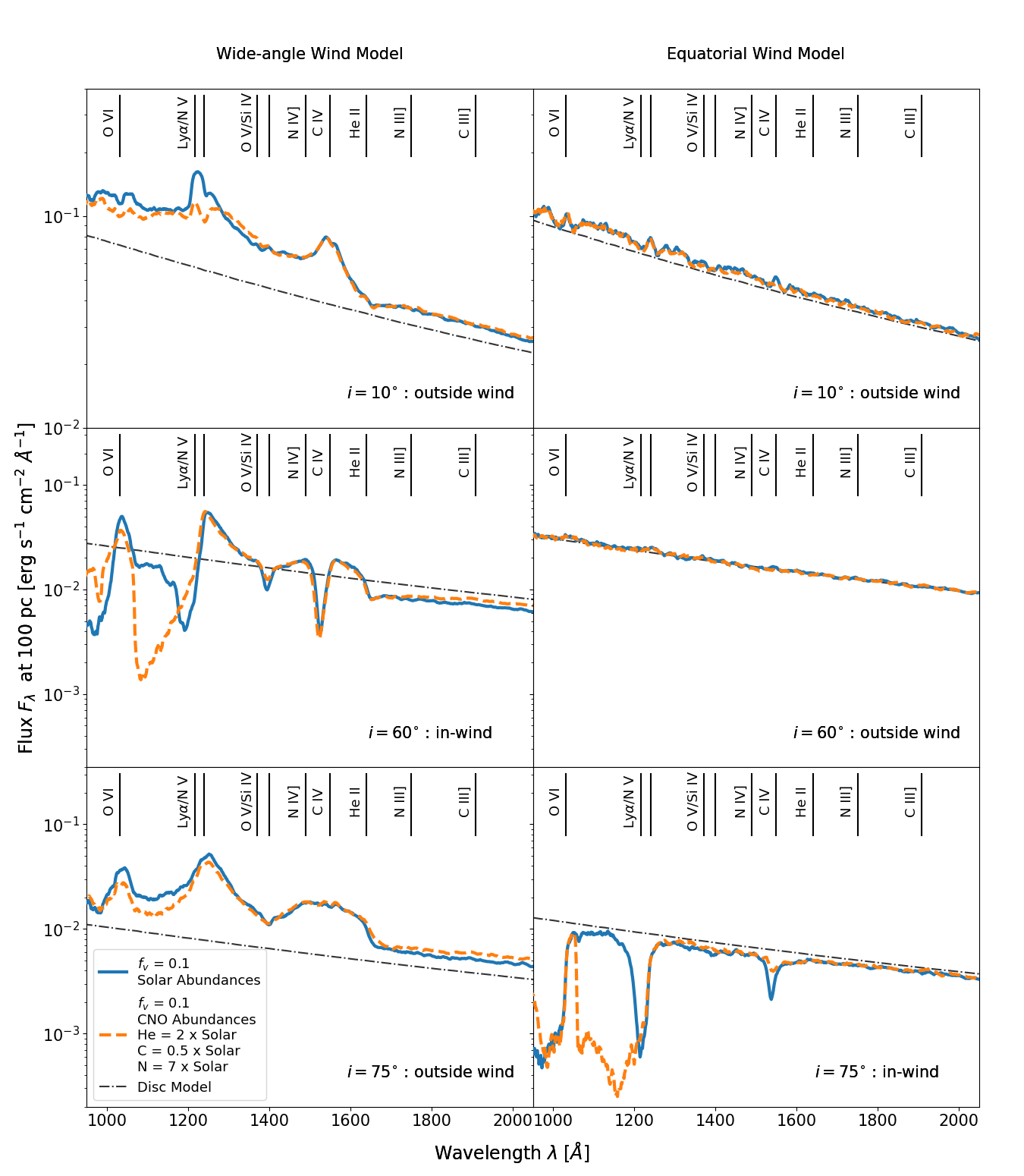}
            \caption{Synthetic UV spectra generated for clumpy winds for our two wind geometries for both Solar and CNO-processed abundances at characteristic low, medium and high inclination angles. In the figure Solar wind spectra are displayed as solid lines, whereas CNO-processed wind spectra are shown with dashed lines. The underlying accretion disc SED (dot-dash lines) for each model are also plotted.}
            \label{fig: model_comparison_clump_abundances}
        \end{figure*}

        Some UV spectra of TDEs present strong Nitrogen emission lines, but weak or undetectable Carbon lines; e.g. iPTF15af and ASASSN14li. This has been attributed to an enhanced Nitrogen to Carbon ratio in the line-forming region, as expected from TDEs associated with stars that have undergone CNO-processing. \citep{Kochanek2016, Yang2017}. We have therefore also carried out some simulations with abundances representing CNO-processed material. In these, we changed the relative abundances of Helium, Carbon and Nitrogen to $(X / X_{\odot})_{^{2}He} \approx 2$, $(X / X_{\odot})_{^{6}C} \approx 0.5$ and $(X / X_{\odot})_{^{7}N} \approx 7$, where $X_{\odot}$ is the relevant solar abundance. These values are based on abundances calculated by \citet{gallegos-garcia2018} for a $\simeq 1.8 M_{\odot}$ star near the terminal main sequence. In Figures~\ref{fig: model_comparison_abund} and \ref{fig: model_comparison_clump_abundances}, we explore the impact of abundance variations by showing synthetic spectra produced by smooth and clumpy winds with solar and CNO-processed abundances, respectively.

        For smooth equatorial winds, the change to CNO-processed abundances results in a slight suppression of the C \textsc{iv} line. The temperature and ionization structure of the wind is roughly the same for both abundance patterns, suggesting that this suppression is due to the change in the abundance of Carbon atoms, rather than a change in the ionization state. This suppression is less obvious in the wide-angle wind model, probably because C \textsc{iv} is more dominant, resulting in the line being more optically thick here. In any case, the effect of the CNO-processed abundances on the N \textsc{v} BAL is much greater. In both wind geometries, this feature becomes both deeper and broader as the density of Nitrogen atoms increases in the wind.
        
        The effect of CNO-processed abundances on clumpy winds is similar, but stronger. Figure \ref{fig: model_comparison_clump_abundances} shows a comparison between the two abundances for a clumping factor $f_{v} = 0.1$. For the equatorial wind, we again see that the C \textsc{iv} absorption has weakened, whereas it remains unchanged for the wide-angle wind. As before, the strongest effect is on the N \textsc{v} BAL, which is once again deeper and broader with CNO-processed abundances.
        
 \section{Discussion} \label{sec: discussion}

    \subsection{UV Lines as Geometry and Orientation Indicators} \label{sec: discussion-geo}
    
        BALs have been now detected in three out of four UV spectra of TDEs. As pointed out by \citet{Hung2019}, this suggests that the appearance of these outflows is less sensitive to viewing angle than that of the outflows in QSOs, where BALs are seen in only $\simeq 20$ per cent of sources \citep[e.g.][]{Knigge2008, Allen2010}. If BALs are observed preferentially for sight lines to the central engine that pass through the outflow, the fraction of systems displaying them is a measure of the wind "covering factor", i.e. the solid angle subtended by the wind, as seen by the central engine. In practice, selection effects will complicate this picture. For example, if bright systems are over-represented in observational samples, fore-shortening, limb-darkening and attenuation by the wind itself will all lead to orientation-dependent incompleteness. Moreover, TDEs are evolving systems, so their outflows and associated BALs are bound to be transient features. Nevertheless, for TDEs observed at similar stages of their eruptions, as is the case for the objects shown in Figure \ref{fig: tde_uv_obs}, the incidence rate of BALs should provide a reasonable initial estimate of the outflow covering fraction.
    
        Our finding that a wide-angle wind model seems to match the observations better than an equatorial one is in line with this picture. The covering factor, $f_{\Omega} = \int_{\theta_{min}}^{\theta_{max}} \sin(\theta)~\text{d}\theta$, of our wide-angle wind model is $f_{\Omega} = 0.52$, whereas that of our QSO-inspired equatorial model is $f_{\Omega} = 0.20$. A wide-angle wind model spanning an even wider range of opening angles is clearly worth exploring in future.
    
        In the context of our particular wide-angle wind model, where $\theta_{min} > 0^\circ$ and $\theta_{max} < 90^\circ$, BELs are seen for both very low and very high inclinations. Do TDEs displaying BELs correspond to extremely face-on systems, extremely edge-on ones, or both? For our fiducial model, the narrower line widths seen for face-on orientations appear to be more in line with the very limited observational data (c.f. Figure \ref{fig: best_model_data_comparison_clump}). On the other hand, extremely face-on orientations are a priori unlikely. Only 6 per cent of randomly oriented systems would be viewed at inclinations $i < 20^\circ$, whereas 42 per cent would be viewed at $i > 65^\circ$. In any case, the detailed line shapes produced by our models should not be over-interpreted: the parameter space of even just wide-angle models is large, and, as discussed in Section \ref{sec:limitations} below, our calculations still have significant limitations that are likely to affect the line profiles. As the sample of TDEs with UV spectroscopy grows, it will be interesting to check for correlations between the incidence of BALs/BELs and other system properties that might depend on orientation \citep[such as the X-ray to optical ratio;][]{Dai2018}.

    \subsection{The Optical Depth of Disc Winds in TDEs}
    
        Several lines of evidence suggest that the outflows of TDEs may be optically thick. For example, the likely presence of P \textsc{v} $\lambda 1118$\footnote{Note that this transition is currently not included in \pythonstop's line list.} in iPTF15af suggests a column density of $N_H > 10^{23}$~cm$^{-2}$ \citep{Blagorodnova_2019}. A similar constraint has been obtained for AT2018zr from the detection of BAL signatures in Balmer and metastable Helium lines \citep{Hung2019}. From a modelling perspective, \citet{roth_what_2018} have shown that the detailed line profile shapes seen in TDEs can be understood if these features are formed in an optically thick outflow. Finally, the reprocessing of radiation produced by the central engine in an optically thick, non-spherical wind is also central to the unification scenario presented by \citet{Dai2018}, which is based on 3-D general relativistic magneto-hydrodynamic simulations of the super-Eddington accretion phase near the peak of TDE flares.
    
\begin{figure*}
            \centering
            \includegraphics[scale=0.65]{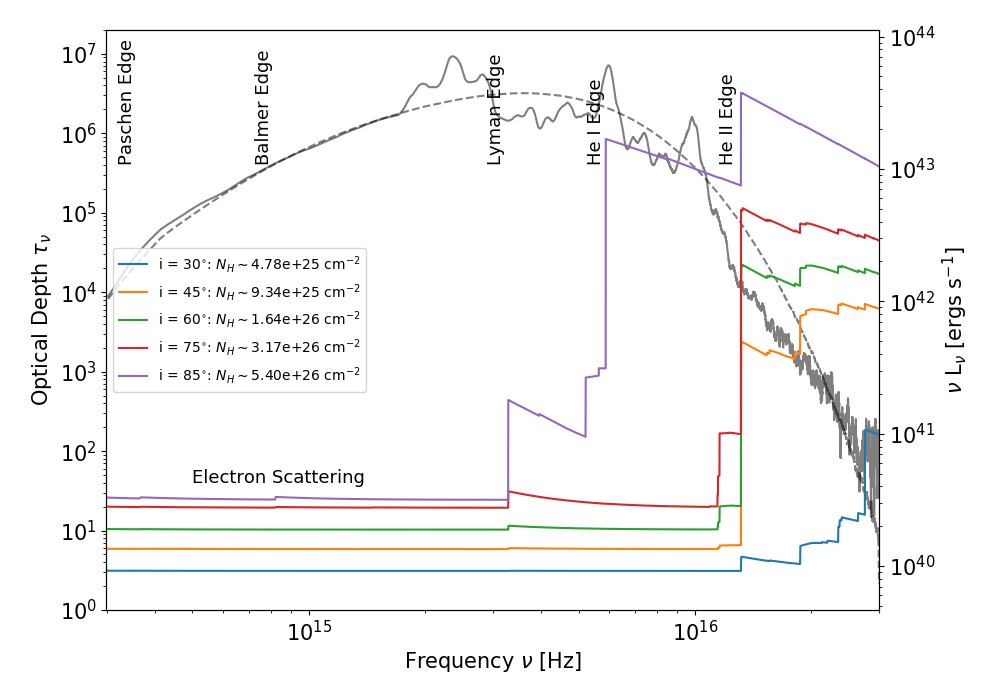}
            \caption{The optical depth as a function of frequency for the fiducial model shown in Figure \ref{fig: best_model_data_comparison_clump} for various sight lines. Also labelled on the plot is the Hydrogen column density $N_{H}$ of each sight line. Marked by the black dashed line is the angle averaged accretion disc SED for the model and shown by the solid black line is the angle averaged emergent flux for the model. Major absorption edges are marked at the top of the figure.}
            \label{fig: polar_optical_depth}
        \end{figure*}

        How optically thick is our fiducial clumpy wide-angle wind model? Figure \ref{fig: polar_optical_depth} shows the continuum (free-free, bound-free and  electron scattering) optical depth through this model, as a function of frequency, for several lines of sight. Starting from a point at the inner disc edge, each sight line here runs radially outward, making an angle $i$ with the normal to the disc plane. At low frequencies, electron scattering is the dominant source of opacity. The electron scattering optical depth is $\tau_{es} > 1$ for all of the sight lines shown ($30^{\circ} \leq i \leq 85^\circ$). For $i = 30^\circ$, the optical depth is moderate, $\tau_{es} \simeq 3$, but it increases monotonically with inclination, reaching $\tau_{es} \simeq 30$ for $i = 85^\circ$. This increase continues even for sight lines that exit the wind cone, i.e. $i > \theta_{max}$, because the shorter path through the wind for these sight lines is more than compensated by the higher densities found at the base of the wind. On average, each photon in our fiducial wide-angle wind models undergoes $\simeq\!\! 100$ electron scatters, with a few photons scattering upwards of 2000 times.
    
        At higher frequencies, the opacities are dominated by photo-ionization. In particular, the He {\sc ii} edge at 54.4 eV (228~\AA) is highly optically thick for essentially all inclinations. The Lyman edge at 13.6 eV (912~\AA) and the He {\sc i} edge at 24.6 eV (504~\AA) only start to make a noticeable contribution to the opacity for high-inclination lines of sight that pass through the dense, less ionized wind base. Figure \ref{fig: polar_optical_depth} also includes the angle-averaged input disc SED and the angle-averaged emergent flux of the model for comparison. This shows that the contribution of He {\sc ii} ionizing photons, which will be almost completely reprocessed by the wind, is not dominant, but still significant in our fiducial model. The majority of the reprocessed emission in the model is emitted at optical wavelengths and at high inclination angles. The column densities along the sight lines shown in Figure \ref{fig: polar_optical_depth} range from $N_H \simeq 5\times 10^{25}$~cm$^{-2}$ (at $i=30^\circ$) to $N_H \simeq 5\times 10^{26}$~cm$^{-2}$ (at $i = 85^\circ$). These values are compatible with the lower limits suggested by \citet{Blagorodnova_2019} and \citet{Hung2019}. 

    \subsection{Limitations and Future Work} \label{sec:limitations}
    
        As is the case for any computational model, our calculations have several limitations. First, \python is a time-independent code, so we implicitly assume that the flow is in a steady-state on time-scales short to the time-scales over which the TDE luminosity and SED evolve significantly. Perhaps most importantly, however, \python is currently a purely Newtonian code. In particular, this means we neglect general relativistic effects on the photon paths in our simulations. Whilst the full random walk of photons is simulated, photons travel in straight lines between interaction points throughout the computational domain. There is also no concept of an event horizon. As a simple test of this approximation, we repeated the simulation for our fiducial wide-angle wind, but this time removed all photons which intersected a sphere of radius $R_{\text{ISCO}}$, centred on the BH. We found that the ionization state and emergent spectra changed only marginally relative to the fiducial model. We also do not take into account the full detailed effects of special relativity, which will have an effect on the Doppler shifts experienced by photons and hence on the emergent line profiles. However, we do not expect this to affect the overall ionization state of the flow, nor the qualitative appearance of the emission and absorption line spectra. Nevertheless, the implementation of relativistic effects into \python is a high priority in our continuing development of the code and future work.
    
        Another clear limitation is our description of the accretion disc as a standard steady state $\alpha$-disc \citep{Shakura1973}. In reality, the inner discs in these systems are dominated by radiation pressure and are almost certainly vertically extended. However, the structure, evolution, and even the stability, of such discs are the subject of intense research in radiation-dominated regimes \citep[e.g.][]{Hirose2009, Jiang_2013, blaes14, Shen_2014}. Additionally, relativistic effects of accretion onto a Kerr black hole is also likely to play a crucial role in TDEs \citep[][]{Balbus2018, Mummery2019a}, particularly at late times \citep{Mummery2019b}. In the absence of a simple, physically realistic description of such discs, our reliance on the standard $\alpha$-disc picture is mainly a practical choice. From the perspective of modelling the outflow, the most important aspect of this approximation may be the angular distribution it imposes on the photons produced by the disc. Due to fore-shortening and limb-darkening, geometrically thin and optically thick discs generate a highly anisotropic radiation field. We adopt the Eddington approximation for limb-darkening, so that the specific intensity produced by the disc scales as $I \propto \cos{i} (1 + \frac{3}{2}\cos{i})$ in our models. This degree of anisotropy is likely to be an overestimate for the vertically extended inner discs in rapidly accreting TDEs. Given that the outflow itself reprocesses, and hence isotropises the disc radiation field, we again do not expect this to change our results qualitatively. However, a more physically realistic description of the accretion discs in TDEs may be needed to enable meaningful \textit{quantitative} modelling of observational data.
    
        Finally, in \pythonstop, only H and He are self-consistently described with full, multi-level model atoms. By contrast, bound-bound transitions in metals are currently treated via a 2-level atom approximation that is reasonable for resonance lines, but not for transitions involving excited and/or meta-stable states. As a result, our simulations are not able to produce semi-forbidden transitions at the moment, such as N \textsc{iii]} or C \textsc{iii]}, which are present in the data and could, in future, be useful outflow diagnostics. We plan to add an approximate way to handle some of these transitions in the near future.

\section{Summary} \label{sec: conclusion}

    Line formation in an accretion disc wind may account for the BEL vs BAL dichotomy observed in the UV spectra of TDEs. In order to test this hypothesis, we have conducted Monte Carlo radiative transfer simulations to create synthetic UV spectra of wind-hosting TDEs, based on a simple biconical disc wind parameterization. Our models cover a wide range of wind geometries and kinematics, and we have also studied the effects of wind clumping and CNO-processed abundances on the spectra. The models presented in this work are available online\footnote{\url{https://github.com/saultyevil/tde_uv_disc_winds_2020}} or upon request. Our main results are are follows:

    \begin{enumerate}
        \item Disc winds are naturally able to produce both BALs and BELs, depending on the sight line of the observer. Sight lines which look through the wind cone preferentially produce BALs, whereas other orientations, which do not look through the wind, tend to produce BELs.
        \item Our "best-bet" fiducial model is a clumpy wide-angle accretion disc wind that subtends more than 50 per cent of the sky as seen from the central engine. Such a geometry is consistent with the high covering factor implied by the detection of BALs in three out of four TDEs with UV spectra to date.
        \item Clumping is required to lower the ionization state of the wind and results in an increased abundance of the relevant ionic species. Both absorption and emission lines are broader and stronger in clumpy wind models, relative to the same lines produced by smooth wind models.
        \item The main effects of switching from solar to CNO-processed abundances are a weakening of C \textsc{iv} $\lambda 1550$ features and a strengthening of N \textsc{v} $\lambda 1240$ ones. Both of these effects are due to the change in the abundances of the relevant species, rather than a major change in the ionization state of the wind.
        \item At long wavelengths, the dominant source of continuum opacity in our models is electron scattering, with photons undergoing $\simeq\!\! ~100$ interactions, on average, before escaping. At shorter wavelengths, and especially for sight lines passing through the dense base of the outflow, the dominant source of opacity is photo-ionization. In particular, the He \textsc{ii} edge is highly optically thick for all sight lines in our fiducial model.
        \item The column densities presented by our fiducial model lie in the range $3\times 10^{25}~{\rm{cm^{-2}}} \lesssim N_H \lesssim 3\times 10^{26}{\rm{cm^{-2}}}$, for inclinations between $i=30^\circ$ and  $i=85^\circ$, with higher columns corresponding to higher inclinations. This is consistent with the empirically inferred lower limits of $N_H \gtrsim 10^{24}$ in BAL-hosting TDEs.
    \end{enumerate}
           
\section*{Acknowledgements}

    We thank the anonymous reviewer for their careful reading of the manuscript, and their helpful and insightful feedback. Figures were prepared using \texttt{matplotlib} \citep{Hunter2007}. The authors acknowledge the use of the IRIDIS High Performance Computing Facility, and associated support services at the University of Southampton. The authors acknowledge the use of the \texttt{GNU Science Library} \citep{gsl}. EJP would like to acknowledge financial support from the EPSRC Centre for Doctoral Training in Next Generation Computational Modelling grant EP/L015382/1. JHM acknowledges a Herchel Smith Research Fellowship at Cambridge. NSH and CK acknowledge support from the Science and Technology Facilities Council grant ST/M001326/1. KSL acknowledges partial support for this project by NASA through grant number HST-GO-14637 from the Space Telescope Science Institute, which is operated by AURA, Inc., under NASA contract NAS 5-26555.

%%%%%%%%%%%%%%%%%%%%%%%%%%%%%%%%%%%%%%%%%%%%%%%%%%

%%%%%%%%%%%%%%%%%%%% REFERENCES %%%%%%%%%%%%%%%%%%

\bibliographystyle{mnras}
\bibliography{bib_main}

%%%%%%%%%%%%%%%%%%%%%%%%%%%%%%%%%%%%%%%%%%%%%%%%%%

%%%%%%%%%%%%%%%%% APPENDICES %%%%%%%%%%%%%%%%%%%%%

\appendix

%%%%%%%%%%%%%%%%%%%%%%%%%%%%%%%%%%%%%%%%%%%%%%%%%%

% Don't change these lines
\bsp	% typesetting comment
\label{lastpage}
\end{document}